# Uncertain Centroid based Partitional Clustering of Uncertain Data

Francesco Gullo
Yahoo! Research
Av. Diagonal, 177
08018 Barcelona, Spain
gullo@yahoo-inc.com

Andrea Tagarelli
DEIS Department
University of Calabria
87036 Rende (CS), Italy
tagarelli@deis.unical.it

## ABSTRACT

Clustering uncertain data has emerged as a challenging task in uncertain data management and mining. Thanks to a computational complexity advantage over other clustering paradigms, partitional clustering has been particularly studied and a number of algorithms have been developed. While existing proposals differ mainly in the notions of cluster centroid and clustering objective function, little attention has been given to an analysis of their characteristics and limits. In this work, we theoretically investigate major existing methods of partitional clustering, and alternatively propose a well-founded approach to clustering uncertain data based on a novel notion of cluster centroid. A cluster centroid is seen as an uncertain object defined in terms of a random variable whose realizations are derived based on all deterministic representations of the objects to be clustered. As demonstrated theoretically and experimentally, this allows for better representing a cluster of uncertain objects, thus supporting a consistently improved clustering performance while maintaining comparable efficiency with existing partitional clustering algorithms.

## 1. INTRODUCTION

Uncertainty in data naturally arises from a variety of real-world phenomena, such as implicit randomness in a process of data generation/acquisition, imprecision in physical measurements, and data staling [1]. For instance, sensor measurements may be imprecise at a certain degree due to the presence of various noisy factors (e.g., signal noise, instrumental errors, wireless transmission) [6]. Another example is given by moving objects [19], which continuously change their location so that the exact positional information at a given time can only be estimated when there is a certain latency in communicating the position (i.e., data is inherently obsolete). The biomedical research domain abounds of data inherently affected by uncertainty. As an example, in the context of gene expression microarray data, handling the so-called probe-level uncertainty represents a key aspect that allows for a more expressive data representation and a more accurate processing [15]. Further examples of uncertain data come from distributed applications, privacy preserving data mining, and forecasting or other statistical techniques used to generate data attributes [1].

In general, uncertainty can be considered at *table*, *tuple* or *attribute* level, and is formally specified by *fuzzy models*, *evidence-oriented models*, or *probabilistic models* [18]. This work focuses on data containing attribute-level uncertainty, which is probabilistically modeled. In particular, we are interested in probabilistic representations that use *probability distributions* to describe the likelihood that any object appears at each position in a multidimensional space [4,12–14]. We hereinafter refer to data objects described in terms of probability distributions as *uncertain objects*.

Given a set of data objects, the problem of *clustering* is to discover a number of homogeneous subsets of objects, called *clusters*, which are well-separated from each other [10]. In the context of uncertain data, *clustering uncertain objects* has recently emerged as a very challenging problem aimed at extending the traditional clustering methods (originally conceived to work on deterministic objects) to deal with objects represented in terms of probability distributions.

Partitional approaches to clustering of uncertain objects include the fastest algorithms so far defined, namely *UK-means* [4,14], which is an adaptation of the popular K-means clustering algorithm to the context of uncertain objects, and *MMvar* [8], which exploits a criterion based on the minimization of the variance of cluster mixture models. Both algorithms involve formulations based on two main notions: *cluster centroid*, which is an object/point that summarizes the information of a given cluster, and *cluster compactness* which is based on the assessment of proximity between the uncertain objects assigned to the cluster and the corresponding centroid. These notions represent a key aspect for the efficiency of partitional methods; in fact, different formulations based on pairwise comparisons between the objects in a cluster (e.g., [7]) are inevitably less efficient.

Though quite efficient, both UK-means and MMVar formulations however suffer from some critical weaknesses that limit the effectiveness of such algorithms. In UK-means the centroid of a cluster of uncertain objects is reduced to be a deterministic (i.e., non-uncertain) point that is simply defined as the average of the expected values of the objects belonging to that cluster. As a consequence, the notion of within-cluster uncertainty in UK-means discards any information about the variance of the cluster members, and hence of the cluster itself. It is quite intuitive that equipped

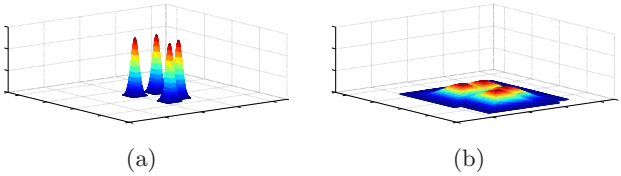

**Figure 1: Uncertain objects with same central tendency: (a) lower-variance, more compact cluster, and (b) higher-variance, less compact cluster**

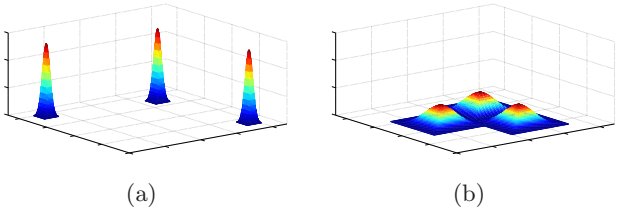

**Figure 2: Uncertain objects with different central tendency: (a) lower-variance, less compact cluster, and (b) higher-variance, more compact cluster**

with such a simplistic definition of cluster centroid, the algorithm can easily fail in distinguishing clusters with the same central tendency but different variance. For instance, consider the situation in Figure 1: since the group of objects in Figure 1-(a) has the same (sum of) expected values as the group in Figure 1-(b), taking into account the variance of these groups is essential to recognize the one with lower variance (Figure 1-(a)) as more compact than the other with higher variance (Figure 1-(b)). As we formally show in Section 4, an analogous issue affects MMVar too. On the other hand, however, considering only the variance to determine cluster compactness may lead to wrong results as well. As an example, consider the objects in Figure 2-(a), which have individual variances smaller than the objects in Figure 2-(b): the latter set of objects clearly represents a cluster more compact than the set in Figure 2-(a); however, a criterion based only on the minimization of the variance of the uncertain objects would mistakenly recognize the group in Figure 2-(a) as better than the group in Figure 2-(b).

In light of the above remarks, a major goal of this work is to address the problem of partitional clustering of uncertain objects in order to improve the accuracy of existing partitional clustering methods, while maintaining high efficiency. Our contributions can be summarized as follows. We provide a deep insight into the UK-means and MMVar formulations, and formally describe the aforementioned issues. We prove that the UK-means and MMVar objective functions differ only by a constant factor, and that no improvement is obtained by employing an objective function that would mix the UK-means cluster compactness criterion and the MMVar cluster centroid definition. Therefore, we propose a formulation to the problem of clustering uncertain objects based on a novel notion of cluster centroid. The proposed centroid, called *U-centroid*, is an uncertain object defined in terms of a random variable whose realizations correspond to all possible deterministic representations deriving from the (deterministic) representations of the uncertain objects to be clustered. We derive the analytical expressions of domain region and probability distribution of the proposed U-centroid. However, as these expressions are in general not analytically computable, we define a cluster compactness criterion that can be efficiently computed according to some closed-form. In particular, after proving that applying the MMVar criterion to the proposed U-centroid is not suited for measuring cluster compactness—it reduces to only consider the variances of the individual uncertain objects, which is not sound as illustrated in Figure 2—we show that minimizing the sum of the distances between each object in a cluster and the corresponding U-centroid fulfils all desired requirements. We propose a local search-based heuristic algorithm, called *U-Centroid-based Partitional Clustering* (UCPC), which searches for a local minimum of the objective function of the proposed formulation by suitably exploiting the aforementioned cluster compactness closed-form expression, thus guaranteeing both effectiveness and efficiency requirements.

We have conducted an extensive experimental evaluation on several data, including datasets with uncertainty generated synthetically as well as real-world collections in which uncertainty is inherently present. The proposed UCPC algorithm outperforms state-of-the-art partitional, density-based and hierarchical algorithms in terms of clustering accuracy (according to external and internal cluster validity criteria). Moreover, from an efficiency viewpoint, UCPC is comparable to the fastest existing partitional methods, i.e., UK-means and MMVar, and can also perform better than pruning-based variants of the basic UK-means algorithm.

The rest of the paper is organized as follows. Section 2 provides background notions and definitions that will be used in this work. Section 3 theoretically shows the weaknesses of UK-means and MMvar. Section 4 discusses our proposal in detail. Section 5 presents experimental settings and results. Section 6 concludes the paper. In Appendix, proof sketches of the main results are finally provided.

## 2. BACKGROUND

Existing research on clustering uncertain objects has focused mainly on adapting traditional clustering algorithms to handle uncertainty. Adaptations have been made for each of the major categories of clustering methods, namely *partitional* (UK-means [4, 14], UK-medoids [7], and MMVar [8]), *hierarchical* (U-AHC [9]), and *density-based* ($\mathcal{F}$DBSCAN [12], $\mathcal{F}$OPTICS [13]). Note that, due to the well-known computational complexity advantages w.r.t. the other clustering paradigms, the partitional clustering methods have attracted more attention, and a number of works have also been devoted to improve the runtime performance of K-means like clustering methods [5, 11, 14, 16, 17].

In the following, we provide formal details on how uncertain objects are represented, and key notions in the two methods mainly under consideration in this work, namely UK-means and MMVar.

### 2.1 Modeling Uncertainty

Uncertain objects are typically represented according to *multivariate* uncertainty models [8], which involve multidimensional domain regions and multivariate pdfs.

DEFINITION 1. *A multivariate uncertain object $o$ is a pair $(\mathcal{R}, f)$, where $\mathcal{R} \subseteq \Re^m$ is the $m$-dimensional domain region* of $o$ and $f : \Re^m \to \Re_0^+$ *is the probability density function of $o$ at each point $\vec{x} \in \Re^m$, such that:*

$$f(\vec{x}) > 0, \ \forall \vec{x} \in \mathcal{R} \quad \text{and} \quad f(\vec{x}) = 0, \ \forall \vec{x} \in \Re^m \setminus \mathcal{R} \quad (1)$$

For arbitrary pdfs of uncertain objects, we assume statistical independence between the actual deterministic representations $\vec{x}, \vec{x}' \in \Re^m$ of any two given uncertain objects $o = (\mathcal{R}, f)$, $o' = (\mathcal{R}', f')$. Formally, $\forall \vec{x}, \vec{x}' \in \Re^m$, it can be assumed that:

$$\Pr(o \equiv \vec{x}, o' \equiv \vec{x}') = \Pr(o \equiv \vec{x}) \ \Pr(o' \equiv \vec{x}') = f(\vec{x}) \ f'(\vec{x}')$$

where "$o \equiv \vec{x}$" denotes the event "*the actual representation of the uncertain object $o$ correspond to the point $\vec{x} \in \Re^m$*".

The expected value ($\vec{\mu}$), second order moment ($\vec{\mu}_2$), and variance ($\vec{\sigma}^2$) vectors of an uncertain object $o = (\mathcal{R}, f)$ are defined as follows:

$$\vec{\mu}(o) = \int_{\vec{x} \in \mathcal{R}} \vec{x} \ f(\vec{x}) \ \mathrm{d}\vec{x} \qquad \vec{\mu}_2(o) = \int_{\vec{x} \in \mathcal{R}} \vec{x}^{\,2} f(\vec{x}) \ \mathrm{d}\vec{x} \quad (2)$$

$$\vec{\sigma}^{\,2}(o) = \int_{\vec{x} \in \mathcal{R}} (\vec{x} - \vec{\mu}(o))^2 f(\vec{x}) \ \mathrm{d}\vec{x} = \vec{\mu}_2(o) - \vec{\mu}^{\,2}(o) \quad (3)$$

The $j$-th component ($j \in [1..m]$) of the $\vec{\mu}, \vec{\mu}_2$ and $\vec{\sigma}^2$ vectors is as follows:

$$\mu_j(o) = \int_{\vec{x} \in \mathcal{R}} x_j \ f(\vec{x}) \ \mathrm{d}\vec{x} \quad (\mu_2)_j(o) = \int_{\vec{x} \in \mathcal{R}} x_j^{\,2} f(\vec{x}) \ \mathrm{d}\vec{x} \quad (4)$$

$$(\sigma^{\,2})_j(o) = \int_{\vec{x} \in \mathcal{R}} (x_j - \mu_j(o))^2 f(\vec{x}) \ \mathrm{d}\vec{x} = (\mu_2)_j(o) - \mu_j^2(o) \quad (5)$$

Moreover, given a vector $\vec{\sigma}^{\,2}$ of variances, the "global" variance expressed in terms of a single numerical value is defined as the sum of variances along each dimension:

$$\sigma^2(o) = \|\vec{\sigma}^{\,2}(o)\|_1 = \sum_{j=1}^m (\sigma^2)_j = \int_{\vec{x} \in \mathcal{R}} \|\vec{x} - \vec{\mu}(o)\|^2 f(\vec{x}) \ \mathrm{d}\vec{x} \quad (6)$$

## 2.2 UK-means

Given a cluster $C$ of uncertain objects, the centroid of $C$ according to the basic UK-means algorithm [4] is a deterministic (i.e., non-uncertain) point $\overline{C}_{UK}$ defined by averaging over the expected values of the objects within $C$:

$$\overline{C}_{UK} = \frac{1}{|C|} \sum_{o \in C} \vec{\mu}(o) \quad (7)$$

Given a candidate clustering $\mathcal{C}$, the basic UK-means minimizes the objective function $\sum_{C \in \mathcal{C}} J_{bUK}(C)$, with $J_{bUK}(C) = \sum_{o \in C} ED_d(o, \overline{C}_{UK})$. $ED_d(\cdot, \cdot)$ denotes the *expected distance* between an uncertain object $o = (\mathcal{R}, f)$ and a point, and is defined as $ED_d(o, \vec{y}) = \int_{\vec{x} \in \mathcal{R}} d(\vec{x}, \vec{y}) \ f(\vec{x}) \ \mathrm{d}\vec{x}$, where $d(\cdot, \cdot)$ is any metric measuring the distance between two $m$-dimensional points. Note that the computation of the integral in $ED_d$ represents a major bottleneck in the execution of the basic UK-means, since in the general case it cannot be computed in a closed-form but it requires an approximation based on a (typically large) set of statistical samples to be drawn from the pdf of the objects. The cost of this integral approximation is not negligible, and hence the overall complexity of the basic UK-means is $\mathcal{O}(I \ S \ k \ n \ m)$, where $n$ is the size of the input set of uncertain objects, $m$ is the dimensionality of the uncertain objects, $k$ is the desired number of clusters,

$S$ is the cardinality of the sample set, and $I$ is the number of iterations for the algorithm convergence.

To speed-up the execution of the basic UK-means, most work has been done on developing *pruning techniques*, whose general goal is to avoid the computation of redundant expected distances between uncertain objects and (candidate) cluster centroids. Major contributions in this regard are proposed in [16] (MinMax-BB algorithm), [11] (VDBiP algorithm), and [17], where the cluster-shift technique is introduced as a general method to further tighten bounds obtained by existing pruning strategies. However, the worst-case complexity of the basic UK-means is not reduced by the pruning techniques. For this purpose, [14] proposes a different approach based on a modification of the formula of the expected distance: denoting simply by $ED(\cdot, \cdot)$ the expected distance $ED_d$ when the metric $d$ is the squared Euclidean norm $\|\cdot\|^2$, [14] shows that:

$$ED(o, \overline{C}_{UK}) = \int_{\vec{x} \in \mathcal{R}} \|\vec{x} - \overline{C}_{UK}\|^2 \ f(\vec{x}) \ \mathrm{d}\vec{x} =$$
$$= ED(o, \vec{\mu}(o)) + \|\overline{C}_{UK} - \vec{\mu}(o)\|^2 \quad (8)$$

Thus, the expected distance between any uncertain object $o$ and centroid $\overline{C}_{UK}$ is equal to the sum of two terms: the first, which is the most expensive one, is given by the expected distance between $o$ and its expected value $\vec{\mu}(o)$, while the second one, which is efficiently computable in $\mathcal{O}(m)$, is equal to the (squared) Euclidean distance between the centroid $\overline{C}_{UK}$ and $\vec{\mu}(o)$. Since the first term does not change during the execution of the algorithm (and hence it can be precomputed for each input object), the algorithm in [14] has an "online" complexity of $\mathcal{O}(I \ k \ n \ m)$. Unless otherwise specified, throughout the rest of this paper we refer to the algorithm in [14] as UK-means and to its cluster compactness criterion as

$$J_{UK}(C) = \sum_{o \in C} ED(o, \overline{C}_{UK}) \quad (9)$$

## 2.3 MMVar

In the MMVar algorithm, the centroid of a cluster $C$ is defined as an uncertain object $\overline{C}_{MM}$ that represents a *mixture model* of $C$:

$$\overline{C}_{MM} = (\overline{\mathcal{R}}_{MM}, \overline{f}_{MM}) \quad (10)$$

where $\overline{\mathcal{R}}_{MM} = \bigcup_{o \in C} \mathcal{R}$, and the pdf $\overline{f}_{MM}(\vec{x})$ is defined as the average $|C|^{-1} \sum_{o \in C} f(\vec{x})$ of the pdfs of the objects within $C$.

The cluster compactness criterion $J_{MM}$ employed by MMVar is simply based on the minimization of the variance of the cluster centroid:

$$J_{MM}(C) = \sigma^2(\overline{C}_{MM}) \quad (11)$$

Analogously to UK-means, the overall objective function to be minimized for a candidate clustering $\mathcal{C}$ is $\sum_{C \in \mathcal{C}} J_{MM}(C)$, and the complexity of the algorithm is $\mathcal{O}(I \ k \ n \ m)$ [8].

## 3. COMPARING UK-MEANS AND MMVAR

*UK-means shortcomings.* As previously discussed, UK-means is characterized by a deterministic definition of cluster centroid, which is simply the average of expected values of the cluster members (7). This implies that UK-means does not explicitly take into account the individual variances of the objects that belong to a cluster. As shown in the following proposition, a major consequence is that two clusters can have the same value of objective function $J_{UK}$ regardless of their respective cluster variance (i.e., sum of variances of the objects that belong to a cluster).

PROPOSITION 1. *Given any two clusters $C$ and $C'$ of uncertain objects, it holds that:*

$$J_{UK}(C) = J_{UK}(C') \;\not\Rightarrow\; \sum_{o \in C} \sigma^2(o) = \sum_{o' \in C'} \sigma^2(o')$$

The above proposition states that the compactness criterion at the basis of UK-means might not discriminate among groups of uncertain objects having different cluster variances. Moreover, it can be straightforwardly derived from the proposition that

$$\sum_{o \in C} \sigma^2(o) \neq \sum_{o' \in C'} \sigma^2(o') \;\not\Rightarrow\; J_{UK}(C) \neq J_{UK}(C')$$

i.e., different values of cluster variance for $C$ and $C'$ do not necessarily force the values $J_{UK}(C)$ and $J_{UK}(C')$ to be different. These aspects may lead to situations like that already illustrated in Figure 1, where the cluster algorithm is unable to recognize a cluster with smaller variance as less uncertain, hence more compact, than a cluster with higher variance.

*Comparing UK-means and MMVar objective functions.* The UK-means weaknesses could in principle be overcome by the MMVar criterion $J_{MM}$, since MMVar centroids involve uncertainty in their representation. Unfortunately, this is not true, as $J_{UK}$ and $J_{MM}$ can be demonstrated to be very close to each other. In particular, as formally shown in the following proposition, the UK-means and MMVar objective functions differ from each other only by a constant factor. This clearly implies that the aforementioned UK-means accuracy issues affect MMVar as well, though the MMVar centroid definition involves uncertainty.

PROPOSITION 2. *Let $C$ be a cluster of $m$-dimensional uncertain objects, where $o = (\mathcal{R}, f)$, $\forall o \in C$. In reference to the functions $J_{UK}$ and $J_{MM}$ defined in (9) and (11), respectively, it holds that $J_{MM}(C) = |C|^{-1} J_{UK}(C)$.*

*Trying to overcome the limitations of UK-means and MMVar.* In the attempt of deriving an alternative objective function that overcomes the limitations shared by UK-means and MMVar, a straightforward solution could be to combine the definition of MMVar centroid with the UK-means cluster compactness criterion, obtaining the following objective function $\widehat{J}(C)$ (by contrast, note that taking the notion of centroid from UK-means while employing the MMVar cluster compactness criterion would be meaningless, since the variance of a deterministic centroid is zero):

$$\widehat{J}(C) = \sum_{o \in C} \widehat{ED}(o, \overline{C}_{MM}) \qquad (12)$$

where $\widehat{ED}$ denotes the (squared) expected distance between any two uncertain objects [7], and is exploited here to compute the distance between any object $o = (\mathcal{R}, f) \in C$ and the centroid (mixture model) $\overline{C}_{MM} = (\overline{\mathcal{R}}_{MM}, \overline{f}_{MM})$ of $C$:

$$\widehat{ED}(o, \overline{C}_{MM}) = \int_{\vec{x} \in \mathcal{R}} \int_{\vec{y} \in \overline{\mathcal{R}}_{MM}} \|\vec{x} - \vec{y}\|^2 f(\vec{x}) \overline{f}_{MM}(\vec{y}) \, d\vec{x} \, d\vec{y} \qquad (13)$$

Unfortunately, adopting such an objective function $\widehat{J}$ is not appropriate yet, as $\widehat{J}$ is in turn proportional to the functions $J_{UK}$ and $J_{MM}$, as shown in the following.

PROPOSITION 3. *Let $C$ be a cluster of $m$-dimensional uncertain objects,, where $o = (\mathcal{R}, f)$, $\forall o \in C$. In reference to the functions $J_{UK}$, $J_{MM}$, and $\widehat{J}$ defined in (9), (11), and (12) respectively, it holds that $\widehat{J}(C) = 2 \, |C| \, J_{MM}(C) = 2 \, J_{UK}(C)$.*

## 4. UNCERTAIN CENTROID BASED PARTITIONAL CLUSTERING

### 4.1 U-centroid

We have demonstrated that the weaknesses of both UK-means and MMVar objective functions cannot be overcome even mixing their basic elements, i.e., the notions of cluster centroid and cluster compactness criterion.

To define a sound objective function for partitional clustering of uncertain objects, we propose here a solution based on a novel notion of cluster centroid. Our key idea is to take into account the random variable whose realizations describe all possible *deterministic representations* of the centroid being defined, in such a way that each of these representations corresponds to the point that minimizes a certain distance function (e.g., the (squared) Euclidean distance) between itself and a set of possible representations of the uncertain objects in the given cluster. More precisely, to define the proposed centroid $\overline{C}$ of a given cluster $C$ of uncertain objects, we consider a real-valued random variable $X_C$, whose observation space is comprised of the events "$\vec{x}$ *is the actual deterministic representation of* $\overline{C}$", $\forall \vec{x} \in \Re^m$. The rationale underlying $X_C$ is as follows: since each object within $C$ has a multiple representation due to its own uncertainty, the centroid of $C$ should have in turn a multiple representation that takes into account the various representations of the objects within $C$; in particular, it should be required that each specific representation of the centroid of $C$ derives from the minimization of a certain distance measure (e.g., Euclidean distance) from a set of points, each corresponding to a possible realization of an uncertain object to be summarized.

Figure 3 illustrates the above concept. Three 2-dimensional uncertain objects forming a cluster are represented along with the cluster centroid—for the sake of simplicity, only the domain regions of the uncertain objects are depicted, as the reasoning being explained holds regardless of a particular pdf. The figure shows that the actual representation of the centroid $\overline{C}$ should change according to the specific points considered for $\mathcal{R}'$, $\mathcal{R}''$, and $\mathcal{R}'''$ as actual representations of the uncertain objects $o'$, $o''$ and $o'''$, respectively. For example, the set of points $\{\vec{x}', \vec{x}'', \vec{x}'''\}$ would lead to the representation of $\overline{C}$ corresponding to the point $\vec{x}$.

In this way, our notion of uncertain centroid, named *U-centroid*, is conceived to gain two main conceptual advantages over existing notions of centroid for uncertain object clusters: 1) it addresses the shortcomings that are typical of a deterministic centroid notion (and hence overcomes a major drawback of the UK-means centroid definition), and

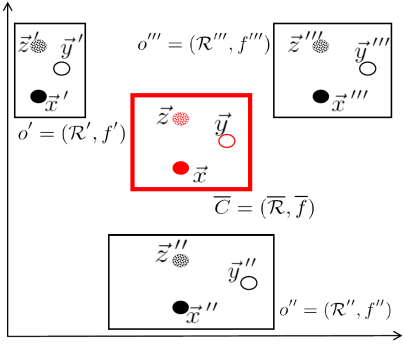

**Figure 3:** Example of uncertain cluster centroid computation based on multiple deterministic representations of uncertain objects

2) it has a clear stochastic meaning. Particularly, the latter is related to an improvement of the notion of centroid in MMVar, as U-centroid is defined in terms of a random variable that describes the outcomes of a clear set of probability events whereas MMVar centroid is a mixture model defined by averaging the pdfs of the objects in a cluster.

In the following, we formally define the proposed notion of U-centroid, as an uncertain object $\overline{C} = (\overline{\mathcal{R}}, \overline{f})$ defined in terms of the random variable $X_C$. Note that $\overline{f}$ is a pdf that expresses the probability of each realization of $X_C$, i.e., $\overline{f}(\vec{x}) = \Pr(\overline{C} \equiv \vec{x}) = \Pr(X_C = \vec{x}), \forall \vec{x} \in \overline{\mathcal{R}}$.

THEOREM 1. *Given a cluster $C = \{o_1, \ldots, o_{|C|}\}$ of $m$-dimensional uncertain objects, where $o_i = (\mathcal{R}_i, f_i)$ and $\mathcal{R}_i = \left[\ell_i^{(1)}, u_i^{(1)}\right] \times \cdots \times \left[\ell_i^{(m)}, u_i^{(m)}\right], \forall i \in [1..|C|]$, let $\overline{C} = (\overline{\mathcal{R}}, \overline{f})$ be the U-centroid of $C$ defined by employing the squared Euclidean norm as distance to be minimized. It holds that:*

$$\overline{f}(\vec{x}) = \int_{\vec{x}_1 \in \mathcal{R}_1} \cdots \int_{\vec{x}_{|C|} \in \mathcal{R}_{|C|}} \mathrm{I}\left[\vec{x} = \frac{1}{|C|} \sum_{i=1}^{|C|} \vec{x}_i\right] \prod_{i=1}^{|C|} f_i(\vec{x}_i) \mathrm{d}\vec{x}_1 \cdots \mathrm{d}\vec{x}_{|C|}$$

$$\overline{\mathcal{R}} = \left[\frac{1}{|C|}\sum_{i=1}^{|C|}\ell_i^{(1)}, \frac{1}{|C|}\sum_{i=1}^{|C|}u_i^{(1)}\right] \times \cdots \times \left[\frac{1}{|C|}\sum_{i=1}^{|C|}\ell_i^{(m)}, \frac{1}{|C|}\sum_{i=1}^{|C|}u_i^{(m)}\right]$$

*where $\mathrm{I}[A]$ is the indicator function, which is 1 when the event A occurs, 0 otherwise.*

It can be straightforwardly derived that $\overline{f}$ and $\overline{\mathcal{R}}$ satisfy the conditions reported in (1), which make any U-centroid an uncertain object according to Definition 1.

## 4.2 U-centroid-based Cluster Compactness

In general, the pdf $\overline{f}$ of any U-centroid reported in Theorem 1 cannot be computed analytically. Hence, we define a cluster compactness criterion based on the notion of U-centroid that does not require to explicitly compute $\overline{f}$.

### 4.2.1 Minimizing the U-centroid Variance

As an U-centroid is an uncertain object, an intuitive definition of a U-centroid-based cluster compactness would be to exploit the same approach as MMVar, i.e., the minimization of the variance of U-centroids. Like in MMVar, a major advantage of this choice would lie in the capability of exploiting an analytical formula to the variance computation. Given a cluster $C$ of uncertain objects, this formula would allow for efficiently computing the variance of the U-centroid of any new cluster $C \cup \{o\}$ (obtained by adding an uncertain object to $C$) or $C \setminus \{o\}$ (obtained by removing an uncertain object from $C$) linearly with the number $m$ of dimensions.

The following theorem shows that the variance of the U-centroid of a cluster $C$ is equal to the average of the variances of the individual uncertain objects within $C$. Though the resulting expression is easy and fast to compute, we prove that however minimizing the variance of the U-centroid is not appropriate to build compact clusters of uncertain objects. In fact, measuring cluster compactness according only to the variances of the individual uncertain objects may lead to wrong results as in general it does not take into account the distances among data objects (cf. Figure 2).

THEOREM 2. *Given a cluster $C = \{o_1, \ldots, o_{|C|}\}$ of $m$-dimensional uncertain objects, where $o_i = (\mathcal{R}_i, f_i), \forall i \in [1..|C|]$, let $\overline{C} = (\overline{\mathcal{R}}, \overline{f})$ be the U-centroid of $C$ defined according to Theorem 1. It holds that $\sigma^2(\overline{C}) = |C|^{-2} \sum_{i=1}^{|C|} \sigma^2(o_i)$.*

### 4.2.2 Minimizing the Expected Distance between Uncertain Objects and U-centroids

After proving that the variance of U-centroid is not effective to measure the cluster compactness, we focus here on a different U-centroid-based criterion, which consists in the minimization of the sum of expected distances between the objects in a cluster and the U-centroid of that cluster:

$$J(C) = \sum_{o \in C} \widehat{ED}(o, \overline{C}) \quad (14)$$

Again, $\widehat{ED}$ denotes the (squared) expected distance between any two uncertain objects (cf. (13)).

In the following theorem, we show that the proposed objective function $J$ overcomes the limitations of all cluster compactness criteria previously discussed in the paper. Specifically, $J$ directly exploits the sum of the variances of the individual objects in a cluster, and hence it explicitly solves the major UK-means/MMVar issue (cf. Section 4 and Figure 1); in addition, $J$ takes into account the sum of expected values of objects as well, which allows for overcoming the issue arising from the criterion based on the minimization of the variance of $\overline{C}$ only (cf. Section 4.2.1 and Figure 2).

THEOREM 3. *Let $C = \{o_1, \ldots, o_{|C|}\}$ be a cluster of $m$-dimensional uncertain objects, where $o_i = (\mathcal{R}_i, f_i), \forall i \in [1..|C|]$, and $\overline{C} = (\overline{\mathcal{R}}, \overline{f})$ be the U-centroid of $C$ defined according to Theorem 1. In reference to the function $J$ defined in (14), it holds that:*

$$J(C) = \sum_{j=1}^{m} \left( \frac{\Psi_C^{(j)}}{|C|} + \Phi_C^{(j)} - \frac{\Upsilon_C^{(j)}}{|C|} \right) = \frac{1}{|C|} \sum_{i=1}^{|C|} \sigma^2(o_i) + J_{UK}(C)$$

*where $J_{UK}$ is the UK-means objective function (cf. (9)) and*

$$\Psi_C^{(j)} = \sum_{i=1}^{|C|} (\sigma^2)_j(o_i) \qquad \Phi_C^{(j)} = \sum_{i=1}^{|C|} (\mu_2)_j(o_i) \qquad \Upsilon_C^{(j)} = \left(\sum_{i=1}^{|C|} \mu_j(o_i)\right)^2$$

COROLLARY 1. *Let $C$ be a cluster of uncertain objects, and $C^+ = C \cup \{o^+\}$, $C^- = C \setminus \{o^-\}$ be two clusters defined by adding an object $o^+ \notin C$ to $C$ and removing an object*

**Algorithm 1** UCPC

**Input:** A set $\mathcal{D}$ of $m$-dimensional uncertain objects; the number $k$ of output clusters
**Output:** A partition (clustering) $\mathcal{C}$ of $\mathcal{D}$
1: compute $\vec{\mu}(o), \vec{\mu}_2(o), \vec{\sigma}^2(o), \forall o \in \mathcal{D}$
2: $\mathcal{C} \leftarrow initialPartition(\mathcal{D}, k)$
3: compute $\Psi_C^{(j)}, \Phi_C^{(j)}, \Upsilon_C^{(j)}, J(C), \forall j \in [1..m], \forall C \in \mathcal{C}$, according to Theorem 3
4: **repeat**
5:    $V \leftarrow \sum_{C \in \mathcal{C}} J(\mathcal{C})$
6:    **for all** $o \in \mathcal{D}$ **do**
7:       let $C^o \in \mathcal{C}$ be the cluster s.t. $o \in C^o$
8:       $C^* \leftarrow \arg\min_{C \in \mathcal{C}} V - [J(C^o) + J(C)] + [J(C^o \setminus \{o\}) + J(C \cup \{o\})]$
9:       **if** $C^* \neq C^o$ **then**
10:         let $C^+ = C^* \cup \{o\}, C^- = C^o \setminus \{o\}$
11:         $\mathcal{C} \leftarrow \mathcal{C} \setminus \{C^*, C^o\} \cup \{C^+, C^-\}$
12:         replace $\Psi_{C^*}^{(j)}, \Phi_{C^*}^{(j)}, \Upsilon_{C^*}^{(j)}, J(C^*)$ with $\Psi_{C^+}^{(j)}, \Phi_{C^+}^{(j)}, \Upsilon_{C^+}^{(j)}, J(C^+), \forall j \in [1..m]$, according to (15)
13:         replace $\Psi_{C^o}^{(j)}, \Phi_{C^o}^{(j)}, \Upsilon_{C^o}^{(j)}, J(C^o)$ with $\Psi_{C^-}^{(j)}, \Phi_{C^-}^{(j)}, \Upsilon_{C^-}^{(j)}, J(C^-), \forall j \in [1..m]$, according to (16)
14:       **end if**
15:    **end for**
16: **until** no object in $\mathcal{D}$ is relocated

**Table 1: Datasets used in the experiments**

(a) Benchmark datasets

| dataset | obj. | attr. | classes |
|---|---|---|---|
| Iris | 150 | 4 | 3 |
| Wine | 178 | 13 | 3 |
| Glass | 214 | 10 | 6 |
| Ecoli | 327 | 7 | 5 |
| Yeast | 1,484 | 8 | 10 |
| Image | 2,310 | 19 | 7 |
| Abalone | 4,124 | 7 | 17 |
| Letter | 7,648 | 16 | 10 |
| KDDCup99 | 4,000,000 | 42 | 23 |

(b) Real datasets

| dataset | obj. | attr. |
|---|---|---|
| Neuroblastoma | 22,282 | 14 |
| Leukaemia | 22,690 | 21 |

Algorithm 1 shows our proposed heuristic algorithm, called *U-Centroid-based Partitional Clustering* (UCPC). After a preliminary phase in which the expected value, second order moment and variance of each object within the input dataset $\mathcal{D}$ are computed (Line 1), UCPC starts by taking an initial partition of $\mathcal{D}$ (Line 2) (e.g., a random partition). Then it follows an iterative procedure such that, at each step, a new clustering is searched to be better than the one obtained at the previous iteration (Lines 4–16). To profitably exploit Theorem 3 and Corollary 1, the new clustering is formed by looking at all possible relocations of an object from its early cluster to a different cluster: for each object $o \in \mathcal{D}$, the relocation that gives the maximum decrement of the objective function w.r.t. the previous clustering will be considered to form the new clustering.

The proposed UCPC converges to a local optimum of the objective function therein involved, and works linearly with both the size of the input dataset and the dimensionality of the input uncertain objects, as formally shown next.

$o^- \in C$ from $C$, respectively. In reference to the expression $J(C) = \sum_{j=1}^{m} \left( |C|^{-1} \Psi_C^{(j)} + \Phi_C^{(j)} - |C|^{-1} \Upsilon_C^{(j)} \right)$, derived in Theorem 3, it holds that:

$$J(C^+) = \sum_{j=1}^{m} \left( \frac{\Psi_{C^+}^{(j)}}{|C|+1} + \Phi_{C^+}^{(j)} - \frac{\Upsilon_{C^+}^{(j)}}{|C|+1} \right) \quad (15)$$

$$J(C^-) = \sum_{j=1}^{m} \left( \frac{\Psi_{C^-}^{(j)}}{|C|-1} + \Phi_{C^-}^{(j)} - \frac{\Upsilon_{C^-}^{(j)}}{|C|-1} \right) \quad (16)$$

where $\Psi_{C^+}^{(j)} = \Psi_C^{(j)} + (\sigma^2)_j(o^+)$, $\Phi_{C^+}^{(j)} = \Phi_C^{(j)} + (\mu_2)_j(o^+)$, $\Upsilon_{C^+}^{(j)} = \left( \sqrt{\Upsilon_C^{(j)}} + \mu_j(o^+) \right)^2$, $\Psi_{C^-}^{(j)} = \Psi_C^{(j)} - (\sigma^2)_j(o^-)$, $\Phi_{C^-}^{(j)} = \Phi_C^{(j)} - (\mu_2)_j(o^-)$, and $\Upsilon_{C^-}^{(j)} = \left( \sqrt{\Upsilon_C^{(j)}} - \mu_j(o^-) \right)^2$.

It should also be emphasized that Theorem 3 provides a closed-form expression for $J$ that does not require to explicitly compute the pdf $\overline{f}$ of $\overline{C}$. This expression, which is efficiently computable in $\mathcal{O}(|C| \, m)$, is given by a linear combination of the terms $\Psi_C^{(j)}, \Phi_C^{(j)}$ and $\Upsilon_C^{(j)}, \forall j \in [1..m]$. This result puts the basis also for an efficient computation of the objective functions $J(C^+)$ and $J(C^-)$ of any two clusters $C^+$ and $C^-$ defined by adding/removing an uncertain object to/from the original cluster $C$. According to Corollary 1, in fact, this can be done in $\mathcal{O}(m)$, as the terms $\Psi_{C^+}^{(j)}, \Phi_{C^+}^{(j)}$ and $\Upsilon_{C^+}^{(j)}$ (resp. $\Psi_{C^-}^{(j)}, \Phi_{C^-}^{(j)}$ and $\Upsilon_{C^-}^{(j)}$) that compose function $J(C^+)$ (resp. $J(C^-)$) can be computed in constant time given the earlier $\Psi_C^{(j)}, \Phi_C^{(j)}$ and $\Upsilon_C^{(j)}$ terms and the expected value, second order moment and variance of the object to be added/removed to/from $C$ (cf. (15)-(16)).

### 4.3 The UCPC Algorithm

The problem of partitional clustering of uncertain objects can be formulated as $\mathcal{C}^* = \arg\min_{\mathcal{C}} \sum_{C \in \mathcal{C}} J(C)$. As it refers to an NP-hard problem, we define a local search-based heuristic that exploits the results reported in Theorem 3 and Corollary 1 to compute effective and efficient approximations.

PROPOSITION 4. *The UCPC algorithm outlined in Alg. 1 converges to a local minimum of function $\sum_{C \in \mathcal{C}} J(C)$ in a finite number of steps.*

PROPOSITION 5. *Given a set $\mathcal{D}$ of $n$ $m$-dimensional uncertain objects, the number $k$ of output clusters, and denoting by $I$ the number of iterations to convergence, the computational complexity of the UCPC algorithm (Alg. 1) is $\mathcal{O}(I \, k \, n \, m)$.*

According to Proposition 5, the proposed UCPC has a complexity equal to that of the fastest existing partitional methods for clustering uncertain objects, i.e., UK-means and MMVar (cf. Section 2); this result proves a major claim of this work, which concerns the efficiency in solving the problem of partitional clustering uncertain objects.

## 5. EXPERIMENTS

Our experimental evaluation was conducted to assess effectiveness, efficiency, and scalability of the proposed UCPC algorithm. For the effectiveness and efficiency evaluations, we used eight benchmark datasets (where the uncertainty was synthetically generated) available from [2] and two real datasets (which originally contained uncertainty) that describe gene expressions in biological tissues (*microarray analysis*) [3]—Table 1 reports on main characteristics of the datasets. Moreover, specifically for the scalability study, we used a very large dataset (4 million objects, last row of Table 1-(a)), which was employed for the KDD Cup '99 contest and now available from the UCI repository [2].

We comparatively evaluated UCPC with the other partitional algorithms, i.e., UK-means (UKM), UK-medoids (UKmed), and MMVar (MMV), with the density-based algorithms, i.e., $\mathcal{F}$DBSCAN ($\mathcal{F}$DB) and $\mathcal{F}$OPTICS ($\mathcal{F}$OPT), and with the agglomerative hierarchical algorithm UAHC. We also included the UK-means variants, namely the basic UK-means (bUKM) and pruning-based methods MinMax-BB and VDBiP (cf. Section 2.2); however, they were considered in the efficiency evaluation only since they share with UK-means the underlying clustering scheme.

To avoid that clustering results were biased by random chance (due to non-deterministic operations, such as computing initial centroids/medoids/partitions), all accuracy and efficiency measurements for each of the algorithms were averaged over multiple (50) runs.

### 5.1 Assessment Methodology

The quality of clustering solutions was evaluated by means of both external and internal criteria.

External criteria exploit the availability of *reference classifications* to evaluate how well a clustering fits a predefined scheme of known classes. Reference classification is hence intended as a predetermined organization of the data objects into a set of classes; clearly, reference classifications were used only for evaluation purposes, and not during the clustering task. We employed the well-known *F-measure* ($F$), which ranges within $[0, 1]$ such that higher values correspond to better quality results. Denoting with $\widetilde{\mathcal{C}} = \{\widetilde{C}_1, \ldots, \widetilde{C}_{\tilde{k}}\}$ a reference classification and with $\mathcal{C} = \{C_1, \ldots, C_k\}$ a clustering solution, F-measure is defined as:

$$F(\mathcal{C}, \widetilde{\mathcal{C}}) = \frac{1}{|\mathcal{D}|} \sum_{u=1}^{\tilde{k}} |\widetilde{C}_u| \max_{v \in [1..k]} F_{uv}$$

where $F_{uv} = (2\ P_{uv}\ R_{uv})/(P_{uv} + R_{uv})$ such that $P_{uv} = |C_v \cap \widetilde{C}_u|/|C_v|$ and $R_{uv} = |C_v \cap \widetilde{C}_u|/|\widetilde{C}_u|$, for each $v \in [1..k]$ and $u \in [1..\tilde{k}]$.

We also used internal criteria based on *intra-cluster* ($intra(\mathcal{C})$) and *inter-cluster* ($inter(\mathcal{C})$) distances (for a given clustering solution $\mathcal{C}$) which express cluster cohesiveness and cluster separation, respectively:

$$intra(\mathcal{C}) = \frac{1}{|\mathcal{C}|} \sum_{C \in \mathcal{C}} \frac{1}{|C|(|C|-1)} \sum_{\substack{o, o' \in C, \\ o \neq o'}} \widehat{ED}(o, o')$$

$$inter(\mathcal{C}) = \frac{1}{|\mathcal{C}|(|\mathcal{C}|-1)} \sum_{\substack{C, C' \in \mathcal{C}, \\ C \neq C'}} \frac{1}{|C| \times |C'|} \sum_{o \in C} \sum_{o' \in C'} \widehat{ED}(o, o')$$

Such distance values are finally combined into a single value $Q(\mathcal{C}) = inter(\mathcal{C}) - intra(\mathcal{C})$, such that the lower $intra(\mathcal{C})$ or the higher $inter(\mathcal{C})$, the better the clustering quality $Q(\mathcal{C})$. Since $intra$ and $inter$ values were normalized within $[0, 1]$, $Q$ ranges within $[-1, 1]$.

*Uncertainty generation.* We synthetically generated uncertainty in benchmark datasets, as they originally contain deterministic values; conversely, this was not necessary for real microarray datasets since they inherently have *probe-level* uncertainty, which can easily be modeled in the form of Normal pdfs according to the *multi-mgMOS* method [15].[1] According to an approach already employed by previous works [4], we developed the following uncertainty generation strategy.

Given a (deterministic) benchmark dataset $\mathcal{D}$, we firstly generated a pdf $f_{\vec{w}}$ for each (deterministic) point $\vec{w}$ within $\mathcal{D}$. In particular, we considered the *Uniform*, *Normal* and *Exponential* pdfs, as they are commonly encountered in real uncertain data scenarios [1]. Every $f_{\vec{w}}$ was defined in such a way that its expected value corresponded exactly to $\vec{w}$ (i.e., $\vec{\mu}(f_{\vec{w}}) = \vec{w}$), whereas all other parameters (such as the width of the intervals of the Uniform pdfs or the standard deviation of the Normal pdfs) were randomly chosen. We exploited the pdfs $f_{\vec{w}}$ to simulate what actually happens in typical real contexts for uncertain data. Thus, we focused on two evaluation cases:

1. the clustering task is performed by considering only the observed (i.e., non-uncertain) representations of the various data objects;

2. the clustering task is performed by involving an uncertainty model.

The ultimate goal was to assess whether the results obtained in Case 2 case are better than those obtained in Case 1.

In Case 1, we generated a *perturbed dataset* $\mathcal{D}'$ from $\mathcal{D}$ by adding to each point $\vec{w} \in \mathcal{D}$ random noise sampled from its assigned pdf $f_{\vec{w}}$ according to the classic Monte Carlo and Markov Chain Monte Carlo methods.[2] As a result, $\mathcal{D}'$ still contains deterministic data. In our evaluation, each of the selected clustering methods was carried out on $\mathcal{D}'$ so that it produced an output clustering solution denoted by $\mathcal{C}'$. A score $F(\mathcal{C}', \widetilde{\mathcal{C}})$ was hence obtained by comparing the output clustering $\mathcal{C}'$ to the reference classification of $\mathcal{D}$ (denoted by $\widetilde{\mathcal{C}}$) by means of the F-measure cluster validity criterion.

In Case 2, when uncertainty is taken into account, we further created an *uncertain dataset* $\mathcal{D}''$ from $\mathcal{D}$ which is the one designed to contain uncertain objects. In particular, for each $\vec{w} \in \mathcal{D}$, we derived an uncertain object $o = (\mathcal{R}, f)$ so that $f = f_{\vec{w}}$, while $\mathcal{R}$ was defined as the region containing most of the area (e.g., 95%) of $f_{\vec{w}}$. Again, we run each of the selected clustering methods on $\mathcal{D}''$ as well, in order to obtain a clustering solution $\mathcal{C}''$ and a score $F(\mathcal{C}'', \widetilde{\mathcal{C}})$.

Finally, we compared the scores obtained in Case 1 and Case 2, respectively, by computing $\Theta(\mathcal{C}', \mathcal{C}'', \widetilde{\mathcal{C}}) = F(\mathcal{C}'', \widetilde{\mathcal{C}}) - F(\mathcal{C}', \widetilde{\mathcal{C}})$; the higher $\Theta$, the better the quality of $\mathcal{C}''$ w.r.t. $\mathcal{C}'$, and, therefore, the better the performance of the clustering method when the uncertainty is taken into account w.r.t. the case where no uncertainty is employed. Note that $\Theta$ ranges within $[-1, 1]$.

### 5.2 Results

#### 5.2.1 Effectiveness

*Accuracy on Benchmark Datasets.* Table 2 shows accuracy results on benchmark datasets for Uniform (U), Normal (N), and Exponential (E) distributions, in terms of both external ($\Theta$) and internal ($Q$) cluster validity criteria (cf. Section 5.1). We also report, for each method, *(i)* the score for

---

[1] We used the Bioconductor package PUMA (*Propagating Uncertainty in Microarray Analysis*) available at http://www.bioinf.manchester.ac.uk/resources/puma/.

[2] We used the SSJ library (www.iro.umontreal.ca/~simardr/ssj/).

616

Table 2: Accuracy results on benchmark datasets: external (F-measure) and internal (Quality) criteria

| data | pdf | F-measure ($\Theta \in [-1, 1]$) | | | | | | | Quality ($Q \in [-1, 1]$) | | | | | | |
|---|---|---|---|---|---|---|---|---|---|---|---|---|---|---|---|
| | | $\mathcal{F}$DB | $\mathcal{F}$OPT | UAHC | UKmed | UKM | MMV | **UCPC** | $\mathcal{F}$DB | $\mathcal{F}$OPT | UAHC | UKmed | UKM | MMV | **UCPC** |
| Iris | U | -.102 | .005 | .002 | .023 | -.062 | .043 | .061 | .197 | .093 | .146 | .148 | .151 | .147 | .145 |
| | N | -.063 | .044 | .017 | .010 | -.010 | .056 | .090 | .238 | .135 | .215 | .194 | .263 | .187 | .197 |
| | E | -.383 | -.174 | -.045 | -.249 | .153 | .161 | | -.004 | .202 | -.001 | .081 | .118 | .692 | .656 |
| Wine | U | -.179 | .174 | .092 | .175 | -.179 | -.011 | -.023 | -.002 | .128 | -.001 | .012 | -.001 | -.001 | -.006 |
| | N | -.185 | .030 | .197 | -.085 | -.184 | .160 | .156 | .022 | .009 | .050 | .042 | -.020 | .124 | .123 |
| | E | -.208 | .006 | -.012 | -.104 | -.208 | -.209 | -.191 | 0 | 0 | 0 | .001 | 0 | .011 | .013 |
| Glass | U | -.298 | .012 | .222 | .084 | .066 | .107 | .423 | -.013 | .001 | .001 | .060 | .001 | .226 | .301 |
| | N | -.040 | -.136 | .132 | -.070 | -.025 | -.009 | .128 | .042 | .006 | .119 | .041 | .057 | .008 | .156 |
| | E | -.334 | -.182 | .131 | .009 | -.231 | .462 | .552 | -.002 | 0 | 0 | .006 | .004 | .140 | .064 |
| Ecoli | U | -.136 | .023 | -.014 | .223 | .199 | .222 | .702 | 0 | .449 | .008 | .187 | .101 | .592 | .648 |
| | N | .061 | .015 | .269 | .045 | .131 | .508 | .533 | .086 | .284 | .088 | .029 | .141 | .151 | .156 |
| | E | -.383 | -.239 | -.129 | -.034 | -.160 | .033 | .003 | 0 | 0 | 0 | .003 | .001 | .187 | .210 |
| Yeast | U | -.085 | .252 | .255 | .315 | .220 | .413 | .642 | 0 | .029 | .001 | .193 | .041 | .566 | .580 |
| | N | .079 | -.001 | .306 | -.035 | .159 | .537 | .620 | .040 | .222 | .150 | .005 | .053 | .253 | .272 |
| | E | -.311 | -.195 | .016 | -.055 | -.098 | .336 | .363 | 0 | 0 | 0 | 0 | 0 | .184 | .160 |
| Image | U | -.283 | -.113 | .046 | .241 | .278 | .071 | .421 | 0 | 0 | 0 | 0 | 0 | .725 | .802 |
| | N | -.251 | -.081 | .127 | -.061 | .122 | .028 | .278 | -.001 | .004 | .130 | .010 | .065 | .004 | .253 |
| | E | -.307 | -.137 | -.020 | .087 | -.024 | .144 | .202 | 0 | 0 | 0 | 0 | 0 | .008 | .119 |
| Abal. | U | -.092 | .291 | .084 | .379 | .120 | .539 | .623 | -.018 | .010 | .060 | .071 | .040 | .226 | .232 |
| | N | .095 | -.039 | .109 | .009 | .034 | .188 | .111 | .086 | .054 | -.030 | .031 | .103 | .057 | .053 |
| | E | -.182 | .315 | .063 | .025 | .080 | .546 | .542 | 0 | 0 | 0 | 0 | 0 | .226 | .283 |
| Letter | U | -.338 | -.201 | .026 | .237 | .008 | .165 | .582 | 0 | 0 | .001 | 0 | 0 | .279 | .297 |
| | N | -.340 | -.203 | .037 | -.039 | -.076 | .127 | .376 | -.022 | .207 | .004 | .357 | .352 | .331 | .305 |
| | E | -.431 | -.294 | .059 | .033 | -.202 | .133 | .153 | 0 | 0 | 0 | 0 | 0 | .147 | .094 |
| avg score | U | -.189 | .055 | .089 | .210 | .081 | .193 | .429 | .021 | .089 | .027 | .084 | .042 | .345 | .375 |
| | N | -.081 | -.046 | .149 | -.028 | .019 | .199 | .287 | .061 | .115 | .091 | .089 | .127 | .139 | .189 |
| | E | -.317 | -.088 | -.008 | -.011 | -.137 | .200 | .223 | -.001 | .025 | 0 | .011 | .015 | .199 | .200 |
| overall avg. score | | -.196 | -.026 | .077 | .057 | -.012 | .198 | .313 | .027 | .076 | .039 | .061 | .061 | .228 | .255 |
| overall avg. gain | | +.509 | +.339 | +.236 | +.256 | +.324 | +.115 | — | +.228 | +.179 | +.216 | +.194 | +.194 | +.027 | — |

Table 3: Accuracy results (Quality) on real datasets

| data | #clust. | Quality ($Q \in [-1, 1]$) | | | | | | |
|---|---|---|---|---|---|---|---|---|
| | | $\mathcal{F}$DB | $\mathcal{F}$OPT | UAHC | UKmed | UKM | MMV | **UCPC** |
| Neuro. | 2 | -.004 | .010 | .917 | .044 | .057 | .592 | .598 |
| | 3 | -.004 | .017 | .670 | .047 | .061 | .600 | .620 |
| | 5 | -.004 | .009 | .847 | .043 | .060 | .678 | .718 |
| | 10 | -.004 | .008 | .607 | .048 | .068 | .098 | .137 |
| | 15 | -.004 | .010 | .578 | .044 | .060 | .675 | .717 |
| | 20 | -.004 | .009 | .487 | .047 | .061 | .582 | .621 |
| | 25 | -.004 | .009 | .465 | .041 | .065 | .596 | .631 |
| | 30 | -.004 | .008 | .466 | .043 | .047 | .532 | .564 |
| Leuk. | 2 | -.018 | .068 | .445 | .221 | .207 | .212 | .224 |
| | 3 | -.018 | .080 | .258 | .256 | .392 | .305 | .352 |
| | 5 | -.018 | .061 | .160 | .245 | .451 | .481 | .537 |
| | 10 | -.018 | .213 | .150 | .238 | .455 | .405 | .451 |
| | 15 | -.018 | .192 | .145 | .246 | .451 | .501 | .544 |
| | 20 | -.018 | .186 | .126 | .213 | .479 | .492 | .528 |
| | 25 | -.018 | .353 | .127 | .215 | .558 | .588 | .620 |
| | 30 | -.018 | .369 | .122 | .213 | .448 | .483 | .512 |
| Neuro. avg score | | -.004 | .010 | .630 | .045 | .060 | .544 | .576 |
| Leuk. avg score | | -.018 | .190 | .192 | .231 | .430 | .433 | .471 |
| over. avg score | | -.011 | .100 | .411 | .138 | .245 | .489 | .523 |
| over. avg gain | | +.534 | +.423 | +.112 | +.385 | +.278 | +.034 | — |

each type of pdf averaged over all datasets (for short, average score), *(ii)* the score averaged over all datasets and pdfs (for short, overall average score), and *(iii)* the overall average gain of the proposed UCPC computed as the difference between the overall average score of UCPC and the overall average score of any specific competing method.

Looking at the overall average scores and gains, the proposed UCPC was more accurate than any other competing method, in terms of both $\Theta$ and $Q$, with gains up to 0.509 ($\Theta$) and 0.228 ($Q$). This finding was confirmed by the results obtained in terms of single dataset-by-pdf configuration. Indeed, according to $\Theta$, UCPC achieved the best results on 17 out of 24 dataset-by-pdf configurations, while, for additional 5 configurations (i.e., all remaining ones except Wine-Uniform and Wine-Exponential), its gap from the best competing method was negligible (smaller than 0.080). Similarly, considering $Q$, UCPC was the best method on the majority (13) of the dataset-by-pdf configurations and achieved results comparable to the best ones in further 9 configurations: only on two configurations (Wine-Uniform and Ecoli-Normal) its gap from the best method was greater than 0.080.

Finally, we remark that the proposed UCPC generally outperformed its most direct competitors UK-means and MMVar, thus confirming a major claim of this work. Compared to UK-means, UCPC achieved better $\Theta$ results on all 24 dataset-by-pdf configurations (gain up to 0.783), whereas in terms of $Q$, UCPC outperformed UK-means on 19 out of 24 configurations (gain up to 0.802), while achieving negligible gaps (smaller than 0.070) in the remaining 5 configurations. UCPC outperformed MMVar as well, though MMVar results were in general better than those achieved by UK-means. Particularly, in terms of $\Theta$, UCPC was better than MMVar on 19 out of 24 dataset-by-pdf configurations (gain up to 0.480), while being comparable on the remaining 5 configurations (gaps smaller than 0.080).

*Accuracy on Real Datasets.* Table 3 shows accuracy results obtained on Neuroblastoma and Leukaemia, and also summarizes *(i)* the scores on each dataset by averaging over the cluster numbers, and *(ii)* the scores and gains by averaging over all cluster numbers and datasets (for short, overall average score). Due to the unavailability of reference classifications for such datasets, we performed multiple tests by varying the number of clusters and assessed the results based on the internal criterion $Q$ only.

UCPC achieved the best overall average performance, with maximum, average and minimum gains (over all the competing algorithms) of 0.534 (w.r.t. $\mathcal{F}$DBSCAN), 0.294, and 0.034 (w.r.t. MMVar), respectively. Moreover, in terms of average scores, UCPC was the best method on Leukaemia, while outperforming all methods but UAHC on Neuroblastoma; however, the gap from UAHC remained small (0.054). Again, our UCPC generally outperformed UK-means and MMVar. Indeed, UCPC was on average more accurate than UK-means on both datasets, while achieving better results on 14 out of 16 dataset-by-number of clusters configurations, whereas, compared to MMVar, UCPC achieved better results on all 16 dataset-by-number of clusters configurations.

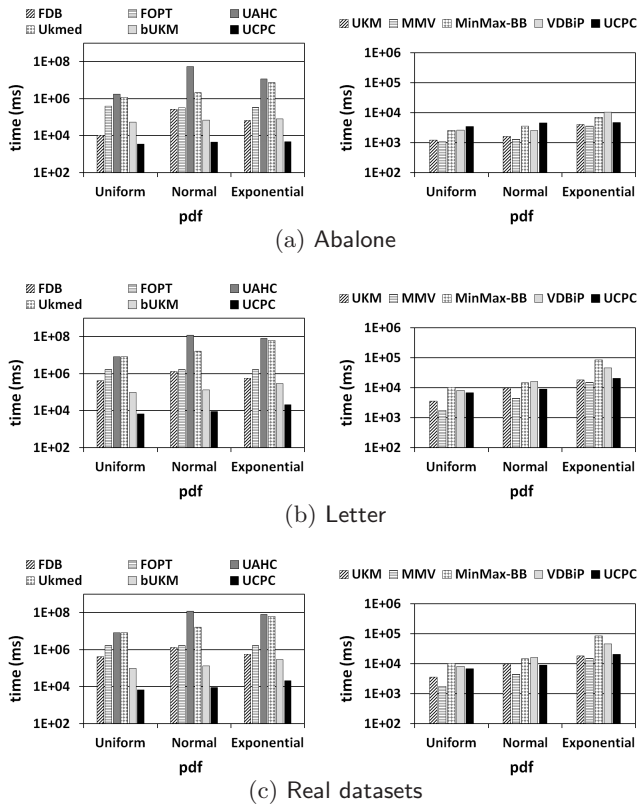

Figure 4: Efficiency results

#### 5.2.2 Efficiency

We also evaluated time performance of our UCPC on both benchmark and real datasets.[3] As previously mentioned, in this evaluation we also included the basic UK-means, MinMax-BB and VDBiP. Actually, in order to possibly strengthen the pruning power of MinMax-BB and VDBiP, they were both coupled with the cluster-shift technique, since it has been demonstrated to have beneficial pruning effects [11, 17]. It is worth emphasizing that the pruning times (i.e., times spent to build and maintain data structures needed for pruning) were discarded from our evaluation, since we chose to focus primarily on the clustering time performance; this also allowed us not to take into consideration R-tree index variants of the Voronoi-diagrams-based pruning (e.g., the RBi algorithm [11]), since R-tree mainly boosts to reduce the pruning time. For an analogous reason, we did not consider the time spent for the off-line stages (i.e., distance pre-computation) required by UK-means and UK-medoids as well. In this respect, we remark that our UCPC does not require any off-line phase.

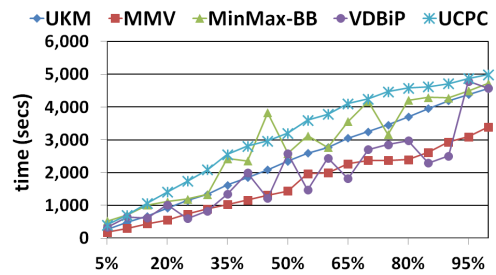

Figure 5: Scalability on the KDD Cup '99 dataset

Figure 4 reports the clustering runtimes (in milliseconds) obtained by the various methods on benchmark and real datasets. Note that, due to space limits of this paper, we present results only for the two largest among the benchmark datasets (excluding KDDCup99); nevertheless, the performance trends we observed on the remaining datasets were roughly similar to those of the datasets here reported. In the figures, results on each dataset are organized in two plots: the left-hand plot contains results obtained by the "slower" algorithms, i.e., UK-medoids, basic UK-means, UAHC, and the density-based algorithms, whereas the right-hand plot contains results obtained by the "faster" MMVar, UK-means and the pruning methods; moreover, to facilitate the comparison with the competing algorithms, each plot also reports the performance of our UCPC.

Looking at the left-hand plots in Figure 4, we observe that UCPC consistently outperformed all the competing algorithms, more specifically: basic UK-means (1 order of magnitude on benchmark datasets, and 2 orders on real datasets), UAHC (3–5 orders), UK-medoids (3–4 orders), $\mathcal{F}$OPTICS (2–3 orders), and $\mathcal{F}$DBSCAN (1–3 orders of magnitude). Concerning the comparison of the "faster" algorithms, UCPC performed very closely to UK-means and MMVar, thus confirming a major claim of this work. Indeed, UCPC achieved times always of the same order of magnitude as UK-means and MMVar. The difference among these three algorithms was very small and negligible in practice. It was also interesting to observe that in most cases, the pruning-based UK-means algorithms (i.e., MinMax-BB and VDBiP) behaved very similarly to each other, and they were always slower than UK-means and MMVar. Note that the worse performance of MinMax-BB and VDBiP w.r.t. UK-means is justified since UK-means does not perform any expected distance computation in the on-line phase. Conversely, the pruning methods significantly improved over the basic UK-means (1 order of magnitude). Moreover, UCPC performed generally better than MinMax-BB and VDBiP and, in some cases, the gap was quite evident (e.g., 1 order of magnitude on the real datasets).

*Scalability.* We carried out a scalability study using the KDD Cup '99 dataset.[4] Figure 5 summarizes the results of this study, for which we varied the dataset size from 5% to 100% and focused on UCPC and the fastest competing algorithms. For each selected subset of the collection, we ensured that all 23 classes were covered by the objects within the subset. Thus, the number of clusters was conveniently fixed to 23 for all the algorithms under consideration.

---

[3]Experiments were conducted on a quad-core platform Intel Pentium IV 3GHz with 4GB memory and running Microsoft WinXP Pro.

[4]For this study, we used the CRESCO HPC system (www.cresco.enea.it), which is integrated into the ENEA-GRID infrastructure. CRESCO is a general purpose system composed by 382 nodes with more than 3300 cores. We executed our experiments on a CentOS 5.5 platform, with Linux 2.6.18 kernel, 64GB memory, 4 Intel(R) Xeon(R) CPU E7330, 2.40GHz quadcore.

As expected, all algorithms (including our UCPC) exhibited linear trends. Particularly, MMVar scaled better than the other algorithms, and UCPC behaved very similarly to UK-means. It was also interesting to observe that the pruning-based UK-means algorithms were subject to some fluctuations, which should be ascribed to a different effect of the pruning on the various dataset portions.

## 6. CONCLUSION

In this paper we defined a novel, well-founded notion of uncertain centroid for clusters of uncertain objects. The proposed notion differs from existing ones in that it represents an uncertain object with a clear stochastic meaning, which conceptually refers to possible deterministic representations of the objects being clustered. Based on this notion, we developed a formulation that overcomes the limitations of existing cluster compactness criteria by taking into account the sum of expected values as well as the sum of the variances of the individual objects in the cluster. Experiments on synthetic and real data have supported our claims of efficiency and, more importantly, improved accuracy in clustering uncertain objects.

## 7. REFERENCES

[1] C. C. Aggarwal. *Managing and Mining Uncertain Data*. Springer, 2009.
[2] A. Asuncion and D. Newman. UCI Machine Learning Repository, http://archive.ics.uci.edu/ml/.
[3] Broad Institute of MIT and Harvard. Cancer program dataset page, http://www.broad.mit.edu/cgi-bin/cancer/datasets.cgi.
[4] M. Chau, R. Cheng, B. Kao, and J. Ng. Uncertain data mining: An example in clustering location data. In *Proc. PAKDD Conf.*, pages 199–204, 2006.
[5] G. Cormode and A. McGregor. Approximation algorithms for clustering uncertain data. In *Proc. ACM PODS Conf.*, pages 191–200, 2008.
[6] A. Deshpande, C. Guestrin, S. Madden, J. M. Hellerstein, and W. Hong. Model-based approximate querying in sensor networks. *The VLDB Journal*, 14(4):417–443, 2005.
[7] F. Gullo, G. Ponti, and A. Tagarelli. Clustering uncertain data via K-medoids. In *Proc. SUM Conf.*, pages 229–242, 2008.
[8] F. Gullo, G. Ponti, and A. Tagarelli. Minimizing the variance of cluster mixture models for clustering uncertain objects. In *Proc. IEEE ICDM Conf.*, pages 839–844, 2010.
[9] F. Gullo, G. Ponti, A. Tagarelli, and S. Greco. A hierarchical algorithm for clustering uncertain data via an information-theoretic approach. In *Proc. IEEE ICDM Conf.*, pages 821–826, 2008.
[10] A. Jain and R. Dubes. *Algorithms for Clustering Data*. Prentice-Hall, 1988.
[11] B. Kao, S. D. Lee, F. K. F. Lee, D. W. L. Cheung, and W. S. Ho. Clustering uncertain data using Voronoi diagrams and R-tree index. *IEEE TKDE*, 22(9):1219–1233, 2010.
[12] H. P. Kriegel and M. Pfeifle. Density-based clustering of uncertain data. In *Proc. ACM KDD Conf.*, pages 672–677, 2005.
[13] H. P. Kriegel and M. Pfeifle. Hierarchical density-based clustering of uncertain data. In *Proc. IEEE ICDM Conf.*, pages 689–692, 2005.
[14] S. D. Lee, B. Kao, and R. Cheng. Reducing UK-means to K-means. In *Proc. IEEE ICDM Workshops*, pages 483–488, 2007.
[15] X. Liu, M. Milo, N. D. Lawrence, and M. Rattray. A tractable probabilistic model for affymetrix probe-level analysis across multiple chips. *Bioinformatics*, 21(18):3637–3644, 2005.
[16] W. K. Ngai, B. Kao, R. Cheng, M. Chau, S. D. Lee, D. W. Cheung, and K. Y. Yip. Metric and trigonometric pruning for clustering of uncertain data in 2D geometric space. *Information Systems*, 36(2):476–497, 2011.
[17] W. K. Ngai, B. Kao, C. K. Chui, R. Cheng, M. Chau, and K. Y. Yip. Efficient clustering of uncertain data. In *Proc. IEEE ICDM Conf.*, pages 436–445, 2006.
[18] A. D. Sarma, O. Benjelloun, A. Y. Halevy, S. U. Nabar, and J. Widom. Representing uncertain data: models, properties, and algorithms. *The VLDB Journal*, 18(5):989–1019, 2009.
[19] G. Trajcevski, O. Wolfson, K. Hinrichs, and S. Chamberlain. Managing uncertainty in moving objects databases. *ACM TODS*, 29(3):463–507, 2004.

## APPENDIX

LEMMA 1. *Given a cluster $C$ of $m$-dimensional uncertain objects, where $o = (\mathcal{R}, f)$, $\forall o \in C$, the function $J_{UK}$ defined in (9) is equal to:*

$$J_{UK}(C) = \sum_{j=1}^{m} \left( \sum_{o \in C} (\mu_2)_j(o) - \frac{1}{|C|} \left( \sum_{o \in C} \mu_j(o) \right)^2 \right)$$

□

PROPOSITION 1. *Given any two clusters $C$ and $C'$ of uncertain objects, it holds that:*

$$J_{UK}(C) = J_{UK}(C') \not\Rightarrow \sum_{o \in C} \sigma^2(o) = \sum_{o' \in C'} \sigma^2(o')$$

PROOF SKETCH. It is sufficient to find a case where $J_{UK}(C) = J_{UK}(C')$ holds and $\sum_{o \in C} \sigma^2(o) = \sum_{o' \in C'} \sigma^2(o')$ does not. To this end, let us assume that: 1) $|C| = |C'|$, 2) $\sum_{j=1}^{m} \sum_{o \in C} (\mu_2)_j(o) = \sum_{j=1}^{m} \sum_{o' \in C'} (\mu_2)_j(o')$, 3) $\sum_{o \in C} \mu_j(o) = \sum_{o' \in C'} \mu_j(o')$, $\forall j \in [1..m]$, and 4) $\sum_{j=1}^{m} \sum_{o \in C} \mu_j^2(o) \neq \sum_{j=1}^{m} \sum_{o' \in C'} \mu_j^2(o')$. Considering assumptions 1)–3) and according to Lemma 1, it follows that:

$$\sum_{o \in C} \mu_j(o) = \sum_{o' \in C'} \mu_j(o'), \ \forall j \in [1..m]$$

$$\Rightarrow \left( \sum_{o \in C} \mu_j(o) \right)^2 = \left( \sum_{o' \in C'} \mu_j(o') \right)^2, \ \forall j \in [1..m]$$

$$\Rightarrow \sum_{j=1}^{m} \left( \sum_{o \in C} \mu_j(o) \right)^2 = \sum_{j=1}^{m} \left( \sum_{o' \in C'} \mu_j(o') \right)^2$$

$$\Rightarrow \sum_{j=1}^{m}\sum_{o\in C}(\mu_2)_j(o) - \frac{1}{|C|}\sum_{j=1}^{m}\left(\sum_{o\in C}\mu_j(o)\right)^2 =$$
$$= \sum_{j=1}^{m}\sum_{o'\in C'}(\mu_2)_j(o') - \frac{1}{|C'|}\sum_{j=1}^{m}\left(\sum_{o'\in C'}\mu_j(o')\right)^2$$
$$\Rightarrow J_{UK}(C) = J_{UK}(C')$$

Similarly, considering also assumption 4), it can be further derived that:

$$\sum_{j=1}^{m}\sum_{o\in C}\mu_j^2(o) \ne \sum_{j=1}^{m}\sum_{o'\in C'}\mu_j^2(o') \Rightarrow \sum_{o\in C}\sigma^2(o) \ne \sum_{o'\in C'}\sigma^2(o')$$
□

LEMMA 2. *Let $C$ be a cluster of uncertain objects, where $o = (\mathcal{R}, f)$, $\forall o \in C$, and $\overline{C}_{MM}$ be the centroid of $C$ employed by MMVar. It holds that:*

$$\vec{\mu}(\overline{C}_{MM}) = \frac{1}{|C|}\sum_{o\in C}\vec{\mu}(o) \qquad \vec{\mu}_2(\overline{C}_{MM}) = \frac{1}{|C|}\sum_{o\in C}\vec{\mu}_2(o)$$
□

PROPOSITION 2. *Let $C$ be a cluster of $m$-dimensional uncertain objects, where $o = (\mathcal{R}, f)$, $\forall o \in C$. In reference to the functions $J_{UK}$ and $J_{MM}$ defined in (9) and (11), respectively, it holds that $J_{MM}(C) = |C|^{-1}J_{UK}(C)$.*

PROOF SKETCH. According to (4) and (6), $J_{MM}(C) = \sigma^2(C)$ can be rewritten as $J_{MM}(C) = \sum_{j=1}^{m}\left((\mu_2)_j(\overline{C}_{MM}) - \mu_j^2(\overline{C}_{MM})\right)$, which, resorting to Lemmas 2 and 1, becomes:

$$J_{MM}(C) = \sum_{j=1}^{m}\left((\mu_2)_j(\overline{C}_{MM}) - \mu_j^2(\overline{C}_{MM})\right) =$$
$$= \frac{1}{|C|}\sum_{j=1}^{m}\left(\sum_{o\in C}(\mu_2)_j(o) - \frac{1}{|C|}\left(\sum_{o\in C}\mu_j(o)\right)^2\right) = \frac{1}{|C|}J_{UK}(C)$$
□

LEMMA 3. *The squared expected distance $\widehat{ED}(o, o')$ between any two $m$-dimensional uncertain objects $o = (\mathcal{R}, f)$ and $o' = (\mathcal{R}', f')$ is equal to $\sum_{j=1}^{m}\left((\mu_2)_j(o) - 2\mu_j(o)\mu_j(o') + (\mu_2)_j(o')\right)$.*
□

PROPOSITION 3. *Let $C$ be a cluster of $m$-dimensional uncertain objects, where $o = (\mathcal{R}, f)$, $\forall o \in C$. In reference to the functions $J_{UK}$, $J_{MM}$, and $\widehat{J}$ defined in (9), (11), and (12) respectively, it holds that $\widehat{J}(C) = 2\,|C|\,J_{MM}(C) = 2\,J_{UK}(C)$*

PROOF SKETCH. According to Lemma 3, function $\widehat{J}$ reported in (12) can be rewritten as:

$$\widehat{J}(C) = \sum_{o\in C}\sum_{j=1}^{m}\left((\mu_2)_j(o) - 2\mu_j(o)\mu_j(\overline{C}_{MM}) + (\mu_2)_j(\overline{C}_{MM})\right) =$$
$$= \sum_{j=1}^{m}\left(\sum_{o\in C}(\mu_2)_j(o) - 2\mu_j(\overline{C}_{MM})\sum_{o\in C}\mu_j(o) + |C|(\mu_2)_j(\overline{C}_{MM})\right)$$

Since $\sum_{o\in C}\mu_j(o) = |C|\mu_j(\overline{C}_{MM})$ and $\sum_{o\in C}(\mu_2)_j(o) = |C| \times (\mu_2)_j(\overline{C}_{MM})$ according to Lemma 2, and $(\sigma^2)_j(o) = (\mu_2)_j(o) - \mu_j^2(o)$ according to (5), the latter expression becomes:

$$\widehat{J}(C) = 2|C|\sum_{j=1}^{m}\left((\mu_2)_j(\overline{C}_{MM}) - \mu_j^2(\overline{C}_{MM})\right) = 2|C|\sum_{j=1}^{m}(\sigma^2)_j(\overline{C}_{MM})$$

By resorting to (6), (11), and Proposition 2, we have finally that:

$$\widehat{J}(C) = 2|C|\sigma^2(\overline{C}_{MM}) = 2|C|J_{MM}(C) = 2J_{UK}(C).$$
□

THEOREM 1. *Given a cluster $C = \{o_1, \ldots, o_{|C|}\}$ of $m$-dimensional uncertain objects, where $o_i = (\mathcal{R}_i, f_i)$ and $\mathcal{R}_i = \left[\ell_i^{(1)}, u_i^{(1)}\right] \times \cdots \times \left[\ell_i^{(m)}, u_i^{(m)}\right]$, $\forall i \in [1..|C|]$, let $\overline{C} = (\overline{\mathcal{R}}, \overline{f})$ be the U-centroid of $C$ defined by employing the squared Euclidean norm as distance to be minimized. It holds that:*

$$\overline{f}(\vec{x}) = \int_{\vec{x}_1\in\mathcal{R}_1}\cdots\int_{\vec{x}_{|C|}\in\mathcal{R}_{|C|}}\mathbb{I}\left[\vec{x} = \frac{1}{|C|}\sum_{i=1}^{|C|}\vec{x}_i\right]\prod_{i=1}^{|C|}f_i(\vec{x}_i)d\vec{x}_1\cdots d\vec{x}_{|C|}$$

$$\overline{\mathcal{R}} = \left[\frac{1}{|C|}\sum_{i=1}^{|C|}\ell_i^{(1)}, \frac{1}{|C|}\sum_{i=1}^{|C|}u_i^{(1)}\right]\times\cdots\times\left[\frac{1}{|C|}\sum_{i=1}^{|C|}\ell_i^{(m)}, \frac{1}{|C|}\sum_{i=1}^{|C|}u_i^{(m)}\right]$$

*where $\mathbb{I}[A]$ is the indicator function, which is 1 when the event $A$ occurs, 0 otherwise.*

PROOF SKETCH. Let us consider sets $\mathcal{S} = \{\{\vec{x}_1, \ldots, \vec{x}_{|C|}\} \mid \vec{x}_1 \in \mathcal{R}_1 \wedge \cdots \wedge \vec{x}_{|C|} \in \mathcal{R}_{|C|}\}$ and $\mathcal{S}_{\vec{x}} = \{S \mid S \in \mathcal{S} \wedge \vec{x} = \arg\min_{\vec{y}\in\Re^m}\sum_{\vec{x}'\in S}d(\vec{y}, \vec{x}')\}$. As $\mathcal{S}$ represents a probability space, it can be exploited for defining a random variable $X_{\mathcal{S}}$, whose realizations $X_{\mathcal{S}} = S$ describe the events that the actual representations of the objects $o_1, \ldots, o_{|C|} \in C$ correspond to the points $\vec{x}_1, \ldots, \vec{x}_{|C|} \in S$, respectively. $X_{\mathcal{S}}$ can be exploited as a conditional variable to derive:

$$f_{X_C}(\vec{x}) = \overline{f}(\vec{x}) = \int_{S\in\mathcal{S}}f_{X_C|X_{\mathcal{S}}=S}(\vec{x}|S)\,f_{X_{\mathcal{S}}}(S)\,dS$$

where $f_{X_{\mathcal{S}}}(S)$ is the pdf of $X_{\mathcal{S}}$. As $f_{X_C|X_{\mathcal{S}}=S}(\vec{x}|S) = 1$ if $S \in \mathcal{S}_{\vec{x}}$, 0 otherwise, as well as $f_{X_{\mathcal{S}}}(S) = \Pr(o_1 \equiv \vec{x}_1 \wedge \cdots \wedge o_{|C|} \equiv \vec{x}_{|C|}) = \prod_{i=1}^{|C|}\Pr(o_i \equiv \vec{x}_i) = \prod_{i=1}^{|C|}f_i(\vec{x}_i)$, $\forall S = \{\vec{x}_1, \ldots, \vec{x}_{|C|}\} \in \mathcal{S}$, it follows that:

$$\overline{f}(\vec{x}) = \int_{S\in\mathcal{S}}f_{X_C|X_{\mathcal{S}}=S}(\vec{x}|S)\,f_{X_{\mathcal{S}}}(S)\,dS =$$
$$= \int_{\vec{x}_1\in\mathcal{R}_1}\cdots\int_{\vec{x}_{|C|}\in\mathcal{R}_{|C|}}\mathbb{I}\left[\vec{x} = \arg\min_{\vec{y}\in\Re^m}\sum_{i=1}^{|C|}d(\vec{y}, \vec{x}_i)\right]\prod_{i=1}^{|C|}f_i(\vec{x}_i)d\vec{x}_1\cdots d\vec{x}_{|C|}$$

By employing the squared Euclidean norm as distance $d$, the optimization problem $\vec{x} = \arg\min_{\vec{y}\in\Re^m}\sum_{i=1}^{|C|}d(\vec{y}, \vec{x}_i)$ becomes $\vec{x} = \arg\min_{\vec{y}\in\Re^m}g(\vec{y})$, where $g(\vec{y}) = \sum_{i=1}^{|C|}\|\vec{y} - \vec{x}_i\|^2$. The solution of such a problem is $\vec{y} = |C|^{-1}\sum_{i=1}^{|C|}\vec{x}_i$. By replacing this into the expression of $\overline{f}(\vec{x})$, we obtain:

$$\overline{f}(\vec{x}) = \int_{\vec{x}_1\in\mathcal{R}_1}\cdots\int_{\vec{x}_{|C|}\in\mathcal{R}_{|C|}}\mathbb{I}\left[\vec{x} = \arg\min_{\vec{y}\in\Re^m}\sum_{i=1}^{|C|}d(\vec{y}, \vec{x}_i)\right]\prod_{i=1}^{|C|}f_i(\vec{x}_i)d\vec{x}_1\cdots d\vec{x}_{|C|}$$
$$= \int_{\vec{x}_1\in\mathcal{R}_1}\cdots\int_{\vec{x}_{|C|}\in\mathcal{R}_{|C|}}\mathbb{I}\left[\vec{x} = \frac{1}{|C|}\sum_{i=1}^{|C|}\vec{x}_i\right]\prod_{i=1}^{|C|}f_i(\vec{x}_i)d\vec{x}_1\cdots d\vec{x}_{|C|}$$

which demonstrates the first statement of the theorem.

The second statement can be proved by observing that all possible representations of the U-centroid $\overline{C}$ can only results by averaging over the possible representations of the uncertain objects within the cluster $C$. □

LEMMA 4. *Let $C = \{o_1, \ldots, o_{|C|}\}$ be a cluster of uncertain objects, where $o_i = (\mathcal{R}_i, f_i)$, $\forall i \in [1..|C|]$, and $\overline{C} = (\overline{\mathcal{R}}, \overline{f})$ be the U-centroid of $C$ defined according to Theorem 1. Given any function $g:\overline{\mathcal{R}} \to \Re$, it holds that:*

$$\int_{\vec{x} \in \overline{\mathcal{R}}} g(\vec{x})\overline{f}(\vec{x})\,\mathrm{d}\vec{x} = \int_{\vec{x}_1 \in \mathcal{R}_1} \cdots \int_{\vec{x}_{|C|} \in \mathcal{R}_{|C|}} g\left(\frac{1}{|C|}\sum_{i=1}^{|C|}\vec{x}_i\right)\prod_{i=1}^{|C|} f_i(\vec{x}_i)\,\mathrm{d}\vec{x}_1 \cdots \mathrm{d}\vec{x}_{|C|}$$

□

LEMMA 5. *Let $C = \{o_1, \ldots, o_{|C|}\}$ be a cluster of uncertain objects, where $o_i = (\mathcal{R}_i, f_i)$, $\forall i \in [1..|C|]$, and $\overline{C} = (\overline{\mathcal{R}}, \overline{f})$ be the U-centroid of $C$ defined according to Theorem 1. It holds that $\vec{\mu}(\overline{C}) = |C|^{-1}\sum_{i=1}^{|C|} \vec{\mu}(o_i)$, and*

$$\vec{\mu}_2(\overline{C}) = \frac{1}{|C|^2}\left(\sum_{i=1}^{|C|}\vec{\mu}_2(o_i) + 2\sum_{i=1}^{|C|-1}\vec{\mu}(o_i)\sum_{i'=i+1}^{|C|}\vec{\mu}(o_{i'})\right)$$

□

THEOREM 2. *Given a cluster $C = \{o_1, \ldots, o_{|C|}\}$ of $m$-dimensional uncertain objects, where $o_i = (\mathcal{R}_i, f_i)$, $\forall i \in [1..|C|]$, let $\overline{C} = (\overline{\mathcal{R}}, \overline{f})$ be the U-centroid of $C$ defined according to Theorem 1. It holds that $\sigma^2(\overline{C}) = |C|^{-2}\sum_{i=1}^{|C|}\sigma^2(o_i)$.*

PROOF SKETCH. From (5) and (6), it follows that $\sigma^2(\overline{C}) = \sum_{j=1}^{m}\left((\mu_2)_j(\overline{C}) - \mu_j^2(\overline{C})\right)$, which, exploiting $\vec{\mu}(\overline{C})$ and $\vec{\mu}_2(\overline{C})$ derived in Lemma 5, can be rewritten as:

$$\sigma^2(\overline{C}) = \sum_{j=1}^{m}\left(\frac{1}{|C|^2}\left(\sum_{i=1}^{|C|}(\mu_2)_j(o_i) + 2\sum_{i=1}^{|C|-1}\mu_j(o_i)\sum_{i'=i+1}^{|C|}\mu_j(o_{i'})\right) + \right.$$
$$\left. - \left(\frac{1}{|C|}\sum_{i=1}^{|C|}\mu_j(o_i)\right)^2\right) =$$
$$= \frac{1}{|C|^2}\sum_{j=1}^{m}\sum_{i=1}^{|C|}\left((\mu_2)_j(o_i) - \mu_j^2(o_i)\right) = \frac{1}{|C|^2}\sum_{i=1}^{|C|}\sigma^2(o_i)$$

□

THEOREM 3. *Let $C = \{o_1, \ldots, o_{|C|}\}$ be a cluster of $m$-dimensional uncertain objects, where $o_i = (\mathcal{R}_i, f_i)$, $\forall i \in [1..|C|]$, and $\overline{C} = (\overline{\mathcal{R}}, \overline{f})$ be the U-centroid of $C$ defined according to Theorem 1. In reference to the function $J$ defined in (14), it holds that:*

$$J(C) = \sum_{j=1}^{m}\left(\frac{\Psi_C^{(j)}}{|C|} + \Phi_C^{(j)} - \frac{\Upsilon_C^{(j)}}{|C|}\right) = \frac{1}{|C|}\sum_{i=1}^{|C|}\sigma^2(o_i) + J_{UK}(C)$$

*where $J_{UK}$ is the UK-means objective function (cf. (9)) and*

$$\Psi_C^{(j)} = \sum_{i=1}^{|C|}(\sigma^2)_j(o_i) \quad \Phi_C^{(j)} = \sum_{i=1}^{|C|}(\mu_2)_j(o_i) \quad \Upsilon_C^{(j)} = \left(\sum_{i=1}^{|C|}\mu_j(o_i)\right)^2$$

PROOF SKETCH. According to Lemma 3, it holds that:

$$J(C) = \sum_{j=1}^{m}\left(\sum_{i=1}^{|C|}(\mu_2)_j(o_i) - 2\,\mu_j(\overline{C})\sum_{i=1}^{|C|}\mu_j(o_i) + |C|(\mu_2)_j(\overline{C})\right)$$

Also, according to Lemma 5, it results that $\mu_j(\overline{C}) = |C|^{-1} \times \sum_{i=1}^{|C|}\mu_j(o_i)$, and

$$(\mu_2)_j(\overline{C}) = \frac{1}{|C|^2}\left(\sum_{i=1}^{|C|}(\mu_2)_j(o_i) + \left(\sum_{i=1}^{|C|}\mu_j(o_i)\right)^2 - \sum_{i=1}^{|C|}\mu_j^2(o_i)\right)$$

Thus, substituting such expressions of $\mu_j(\overline{C})$ and $(\mu_2)_j(\overline{C})$ into function $J$, we obtain:

$$J(C) = \sum_{j=1}^{m}\left(\sum_{i=1}^{|C|}(\mu_2)_j(o_i) - \frac{2}{|C|}\left(\sum_{i=1}^{|C|}\mu_j(o_i)\right)^2 + \right.$$
$$\left. + \frac{1}{|C|}\left(\sum_{i=1}^{|C|}(\mu_2)_j(o_i) + \left(\sum_{i=1}^{|C|}\mu_j(o_i)\right)^2 - \sum_{i=1}^{|C|}\mu_j^2(o_i)\right)\right) =$$
$$= \sum_{j=1}^{m}\left(\frac{\Psi_C^{(j)}}{|C|} + \Phi_C^{(j)} - \frac{\Upsilon_C^{(j)}}{|C|}\right)$$

which proves the first part of the theorem. The second part can be derived by applying the results from Lemma 2:

$$J(C) = \frac{1}{|C|}\sum_{i=1}^{|C|}\sigma^2(o_i) + |C|\sum_{j=1}^{m}\left(\frac{1}{|C|}\sum_{i=1}^{|C|}(\mu_2)_j(o_i) - \left(\frac{1}{|C|}\sum_{i=1}^{|C|}\mu_j(o_i)\right)^2\right)$$
$$= \frac{1}{|C|}\sum_{i=1}^{|C|}\sigma^2(o_i) + |C|\sigma^2(\overline{C}_{MM}) = \frac{1}{|C|}\sum_{i=1}^{|C|}\sigma^2(o_i) + J_{UK}(C)$$

□

PROPOSITION 4. *The UCPC algorithm outlined in Alg. 1 converges to a local minimum of function $\sum_{C \in \mathcal{C}} J(C)$ in a finite number of steps.*

PROOF SKETCH. Let us denote by $V^{(h)}$ the value $\sum_{C \in \mathcal{C}^{(h)}} J(C)$, where $\mathcal{C}^{(h)}$ is the clustering computed at the $h$-th iteration of UCPC. To prove the proposition, it is sufficient to show that $V^{(h)} \leq V^{(h-1)}$ at each iteration $h > 1$, as the function $\sum_{C \in \mathcal{C}} J(C)$ is bounded below. This is true as at each step of the algorithm the optimal move of objects to clusters is performed. □

PROPOSITION 5. *Given a set $\mathcal{D}$ of $n$ $m$-dimensional uncertain objects, the number $k$ of output clusters, and denoting by $I$ the number of iterations to convergence, the computational complexity of the UCPC algorithm (Alg. 1) is $\mathcal{O}(I\ k\ n\ m)$.*

PROOF SKETCH. The initialization (offline) phase (Lines 1-3) takes $\mathcal{O}(k\ n\ m)$, as well as the main cycle (Lines 4-16), thanks to the formulas derived in Corollary 1; this leads to an overall complexity of $\mathcal{O}(I\ k\ n\ m)$. □